\RequirePackage{fix-cm}
\documentclass[twocolumn]{svjour3}          
\smartqed  
\usepackage{graphicx}
\begin{document}

\title{Shape-resonant superconductivity in nanofilms: from weak to strong coupling}


\author{Marco Cariglia \and Alfredo Vargas-Paredes \and Mauro M. Doria \and Antonio Bianconi  \and Milorad V. Milo\v{s}evi\'c \and
        Andrea Perali 
}


\institute{M. Cariglia \at
Departamento de F\'isica, Universidade Federal de Ouro Preto, 35400-000 Ouro Preto Minas Gerais, Brazil and School of Pharmacy, Physics Unit, University of Camerino, 62032 - Camerino, Italy\\
\and
A. A.Vargas-Paredes \at
School of Pharmacy, Physics Unit, University of Camerino, 62032 - Camerino, Italy and Departement Fysica, Universiteit Antwerpen,
Groenenborgerlaan 171, B-2020 Antwerpen, Belgium\\
\and
M. M. Doria \at
Instituto de F\'isica, Universidade Federal do Rio de Janeiro, 21941-972 Rio de Janeiro, Brazil and School of Pharmacy, Physics Unit, University of Camerino, 62032 - Camerino, Italy\\
\and
A. Bianconi \at
Rome International Center for Materials Science Superstripes (RICMASS), Via dei Sabelli 119A, I-00185 Rome, Italy\\
\and
M. V. Mil\v{o}sevi\'c \at
Departement Fysica, Universiteit Antwerpen,
Groenenborgerlaan 171, B-2020 Antwerpen, Belgium\\
\and
A. Perali \at
School of Pharmacy, Physics Unit, University of Camerino, 62032 Camerino and INFN sezione di Perugia, Italy
\email{andrea.perali@unicam.it}\\
}

\date{Received: date / Accepted: date}

\maketitle

\begin{abstract}
Ultrathin superconductors of different materials
are becoming a powerful platform to find mechanisms for enhancement of
superconductivity, exploiting shape resonances in different 
superconducting properties. Here we evaluate the superconducting gap and its
spatial profile, the multiple gap components, and the chemical potential,
of generic superconducting nanofilms, considering the pairing attraction
and its energy scale as tunable parameters, 
from weak to strong coupling, at fixed electron density. Superconducting
properties are evaluated at mean field level as a function of the
thickness of the nanofilm, in order to characterize 
the shape resonances
in the superconducting gap. We find that the most pronounced shape resonances
are generated for weakly coupled superconductors, while approaching
the strong coupling regime the shape resonances are rounded by a mixing
of the subbands due to the large energy gaps extending over 
large energy scales. 
Finally, we find that the spatial profile, transverse to the nanofilm, 
of the superconducting gap
acquires a flat behavior in the shape resonance region, indicating that
a robust and uniform multigap superconducting state can arise at resonance.
\keywords{Shape Resonance \and Ultrathin Superconductivity \and Lifshitz Transitions \and BCS-BEC crossover.}
\PACS{74.20.-z \and 74.20.Fg \and 74.78.-w}
\end{abstract}

\section{Introduction}
\label{intro}

Superconductivity in strongly confined systems at the nano or atomic scale
is attracting a growing interest after the recent observation of a sizable
enhancement of the critical temperature in superconducting FeSe systems
when reduced to monolayers \cite{JinFeng} and the observation of 
superconductivity above 5 K in graphene doped with Lithium \cite{Cao}.
The multiband nature of the superconductivity
in doped FeSe can also lead to amplifications of the superconducting parameters
when the chemical potential crosses a Lifshitz transition \cite{Shi2016}
as well as to BCS-BEC 
crossover phenomena \cite{PeraliPRL2005,Palestini2012} 
in a multigap configuration \cite{Kanigel,Terashima,Okazaki,Gui2014}. 
Furthermore, since 2004 the observation of shape resonances in superconducting
metallic nanofilms of Pb \cite{Guo2004,Eom2006,Qin2009} and first
evidences of shape resonances in the superconducting
critical temperature in metallic
nanowires of Sn and Al \cite{Shanenko2006,Altomarebook,Shanenko_etal2008}
clearly established the importance of the
interplay between quantum size effects and superconductivity
when the lateral dimensions of the system are reduced to the order
of the interparticle distance or
the pair correlation length \cite{Blatt,Perali96}.
It is important to note that current experiments
on nanofilms reporting shape resonances in the gap and in the
critical temperature \cite{Guo2004,Eom2006,Qin2009} 
do not find evidences for multigap superconductivity
close to the shape resonances. In this paper we identify the parameter
regime in which future experiments should directly detect multiple gaps.
Strong enhancement of superconductivity has been also predicted 
and observed when all the lateral dimensions of a bulk superconductor
are reduced to the nanoscale, as in nanoparticles, nanoclusters,
and nanocubes \cite{Garcia-Garcia2008,Garcia-Garcia2010,Garcia-Garcia2014a,Garcia-Garcia2014b}.
Predictions of large amplifications
of the superconducting critical temperature and of multigap BCS-BEC
crossover phenomena point toward superstripes, i.e. a system of periodic
stripes organized in a superlattice, as an ideal candidate system
to control and enhance superconductivity 
at the nanoscale \cite{Bianconi97,Bianconi98,BianconiNP2013}.
Motivated by the fact that many different bulk superconducting
materials can be used as a starting system to realize nanostructures,
for instance by nanosculpting lithography \cite{Fretto2013},
here we investigate theoretically the nature of the superconducting
shape resonances in metallic nanofilms, tuning the parameters of the
pairing interaction from weak to strong coupling, and considering
different values of the energy scale of the pairing.
For a review on theory and experiments discussing the 
multiband and multigap physics of superconducting nanofilms, see 
Refs. \cite{Shanenko_etal2015,MMAP2015}.
The shape resonances in the superconducting gaps at zero temperature
are characterized in terms of the amplification with respect
to the bulk value of the gap and the width of the resonance, where formation
of a multicondensate with multiple gaps can be observed. 
We will show that the amplification
is controlled by the pairing strength, while the width of the
resonance depends on the energy cutoff of the pairing interaction.
Note that recently a heterostructure of superconductors and
insulating barriers has been proposed to generate multigap superconductivity
also outside the shape resonance region \cite{Doria2016}.
The chemical potential renormalization at 
fixed density is also explored, which is important when the system is
close to a shape resonance and the gap becomes a large energy scale with
respect to the distance of the chemical potential to the
Lifshitz transitions \cite{Innocenti2010,PeraliSUST2012}. 
In this situation, a mixture of BCS-like
and crossover BCS-BEC pairs is realized 
\cite{Chen2012,Shanenko2012,Guidini2016}, providing
the best condition to stabilize the detrimental superconducting 
fluctuations \cite{Perali2000} which can be strong in reduced
dimensionality \cite{Marsiglio2015}.
Finally, we investigate
the spatial profile of the superconducting gap parameter and of the
density of electrons. We find that the shape resonant
superconductivity is characterized by a flat behavior of the gap
profile, to be contrasted with a many-peak gap outside resonance.
Resonant superconductivity is therefore the most robust phase of
superconducting nanofilms in the strong quantum confinement regime.

The paper is organized as follows. In Section II we discuss the coupled
mean-field equations for the gaps and the chemical potential, where 
we report the expressions of the gap and density
profiles in the Anderson's approximation to the Bogoliubov-de Gennes
equations. In Section III we report and analyze the numerical results
at $T=0$ for the gaps, the chemical potential and the spatial
profile of the gap and the electronic density. In Section IV we summarize our
results and we conclude with predictions for key experiments 
to detect the effects reported in this paper.

\section{Methods}
\label{methods}

The approach employed in this paper to investigate the ground state
of superconducting nanofilms in the strong quantization
regime is based on a full self-consistent
solution of the coupled gaps and density equations, arising from
the Anderson's prescription \cite{Anderson1959}
to solve the Bogoliubov-de Gennes
equations for non uniform superconductors \cite{deGennes}.
It is the same approach introduced in the seminal paper
by Thompson and Blatt \cite{Blatt}.
Here, we consider a system of electrons confined in a thin metallic slab 
with infinite potential walls. 
In the direction parallel to the film, the electrons have a parabolic
dispersion with an effective mass equal to the bare mass
of the electrons. In the direction perpendicular to the
film the motion of the electrons is quantized, with formation of
discrete single-particle energy levels, as given by the solution of
the uni-dimensional Schr\"odinger equation \cite{Altomarebook,Blatt}.
Hence, the electronic subbands have the following form:
\begin{equation}
\label{dispersion}
\xi_n(\mathbf{k})=\frac{\mid\mathbf{k}\mid^2}{2m}+E_n-\mu \,\,;\,\,\,\,
E_n=\frac{1}{2m}\left(\frac{n\pi}{L}\right)^2,
\end{equation}
where $\mathbf{k}$ is the wave-vector of the electrons parallel to the film,
$m$ is the effective mass,
$\mu$ is the chemical potential, and $E_n$ are the
discrete energies of the subband bottoms.
The index $n$=1,2,... labels the electronic subbands.
For a given chemical potential, the Fermi surface exhibits 
a number of concentric circular 2D Fermi sheets.
The reduced Planck constant ($h/2\pi$) is taken
equal to unity throughout the paper.

The electrons interact via
an effective attraction characterized by an interaction strength $V^0$
and an energy cutoff $\omega_0$.
The effective pairing attraction of the bulk system
is taken in a separable form, as in \cite{Blatt}.
Moreover, because in the case of nanofilms the motion along the $z$-axis
is tightly bound, the bare strengths of the potential that control the
intraband pairings and the interband exchange (Josephson-like) pairing 
between the two subbands
are related by $V^0_{nm}=V^0(1+\frac{1}{2}\delta_{nm})$. This expression
is due to the overlap integral of the single-particle wave-functions,
as arising from the Anderson approximation 
to the full Bogoliubov - de Gennes (BdG) equations
(a detailed comparison between the Anderson approximation and the
exact BdG solution is available in \cite{Tanaka2000}).
Therefore, the quantum confinement in superconducting nanofilms is able
to generate different intraband and pair exchange interactions, but the
partial condensates of each subband are strongly coupled by the pair
exchange terms, being the intraband term only 50\% larger than the pair
exchange. As we will see below, this behavior is at the origin of a not 
too evident multigap structure in single superconducting nanofilms. We note
also that this large pair-exchange interaction will prevent the resonant
condensate to enter the BEC regime at strong coupling \cite{Gui2014}.
The pairing potential can be written as
\begin{eqnarray}
\label{potential}
V_{nm}(\mathbf{k},\mathbf{k}')=-V^0(1+\frac{1}{2}\delta_{nm}) &\Theta&\left(\omega _0 - |\xi_n(\mathbf{k})|\right) \nonumber \\\times &\Theta&(\omega _0 - |\xi_m(\mathbf{k}')|),
\end{eqnarray}
where $V^0$ is the (positive) strength of the attractive potential.
The $\mathbf{k}$-dependence of the (isotropic s-wave) gaps is a consequence 
of the separable
form of the interaction of Eq. (\ref{potential}) and it is given by
\begin{equation}
\label{gaps_k}
\Delta _n(\mathbf{k})=\Delta _n \Theta(\omega _0 - |\xi _n(\mathbf{k})|).
\end{equation}

The coupled mean-field equations for the gaps take the form
originally introduced for two-band superconductors \cite{Suhl59}:
\begin{equation}
\label{gap_eq_1}
\Delta _n(\mathbf{k})=\frac{-1}{\Omega} \sum_{m,\mathbf{k}'}V_{nm}(\mathbf{k},\mathbf{k}')\frac{\Delta _m (\mathbf{k}')}{2\sqrt{\xi _m(\mathbf{k}')^2 +\Delta _m (\mathbf{k}') ^2}},
\end{equation}
$\Omega$ being the surface area of the nanofilm.

In this work the total density of the conduction electrons $n_e$
is fixed at values typical for metals, $n_e=10^{22}/cm^3$, corresponding
to a non interacting Fermi energy in the bulk $E_F=1.7$eV, which will be
our reference value for the chemical potential in the nanofilms in the
limit of large thicknesses. 
At a mean field level at T=0 K the density equation is given by
\begin{equation}
\label{density}
n_e=\frac{2}{\Omega}\sum_{n,\mathbf{k}}v_n(\mathbf{k})^2,
\end{equation}
where the factor 2 is the spin degeneracy of the electrons and
$v_n(\mathbf{k})$ is the BCS weight of the occupied states
\begin{equation}
\label{vk2}
v_n(\mathbf{k})^2=\frac{1}{2}\left[ 1-\frac{\xi_n(\mathbf{k})}{\sqrt{\xi_n(\mathbf{k})^2+\Delta _n(\mathbf{k})^2}}  \right].
\end{equation}

The sums over $\mathbf{k}$ are replaced by two-dimensional integrals over momenta and then by integrals over the energy variable, after introducing the 2D density of states $N_{2D}=m/(2\pi)$. The integrals of Eqs.~(\ref{gap_eq_1}-\ref{density}) can be expressed in a closed form, as shown in Ref. \cite{Altomarebook}.

This self-consistent system of equations for the multiple gaps
and the chemical potential in superconducting nanofilms
has been investigated recently both 
at an analytical and numerical level in Refs. \cite{Valentinis1,Valentinis2},
with their focus on the role of different boundary conditions of the
nanofilms and the continuity of the shape resonances as a function
of thickness.

In this work we will evaluate also the spatial profile 
of the total superconducting gap $\Delta(z)$
and the total density of conduction electrons $n_e(z)$ along the direction
transverse to the nanofilm. Within the Anderson approximation we have,
 
\begin{eqnarray}
\Delta(z)&=&\sum_{n}|\Psi_n(z)|^2\Delta_n\int \frac{dE}{2\sqrt{(E+E_n-\mu)^2+\Delta_n^2}}, \label{gap_z}\\
n_e(z)&=&2\sum_{n}|\Psi_n(z)|^2\int dE v_n^2(E), \label{density_z}
\end{eqnarray}
where $\Psi_n(z)=\sqrt{2/L}\sin(n\pi z/L)$ is the single particle wave-function along the $z$ direction
corresponding to the energy level $E_n$, solution of the Schr\"odinger equation
in the transverse direction with infinite wall potential. 
The extremes of the integral are the same of the coupled
self-consistent gap and density equations, determined by the entering
and exiting of each subband bottom $E_n$ from the Debye energy window. 
In the density profile of Eq. (\ref{density_z}) the
contribution of the free electron density outside the Debye energy window
(hence, with zero gaps) is also included.

\section{Results}
\label{results}

In Figure 1 we report the superconducting gap in the first subband
as a function of the nanofilm thickness for different 3D couplings,
from weak ($\lambda=0.3$) to very strong coupling
($\lambda=2.0$), at fixed energy cutoff of the pairing interaction
($\omega_0=300$ K).
The gaps are normalized to their bulk value, obtained
in the limit of large thickness ($k_FL >>1$).

\begin{figure}
\includegraphics[scale=0.6]{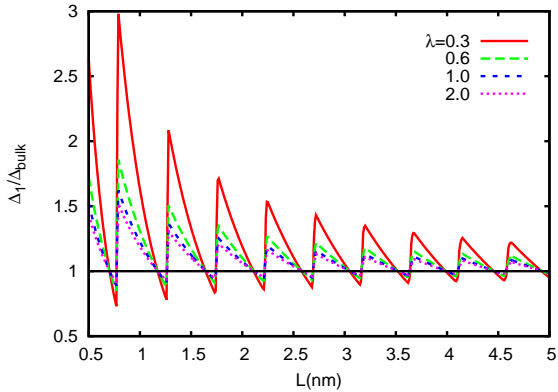}
\caption{Superconducting gap in the first subband $\Delta_1$ as a function of the film thickness $L$ for different couplings. $\Delta_1$ is normalized to the corresponding bulk values of the gap $\Delta_{bulk}$ determined for different couplings (see text).}
\label{fig_1}
\end{figure}

In Table \ref{tab_1} we report the bulk values of the gap for different
couplings, from the weak ($\lambda<0.6$) to the strong ($\lambda>1.0$)
coupling regime. In the third column of Table \ref{tab_1}
we show the amplification
factor $A=\Delta_{max}/\Delta_{bulk}$ of the gap at the shape resonance, taken in the ultrathin regime
at $L=0.8$ nm. For weak coupling the gap amplification is large, while
it approaches values of order unity for stronger couplings.
Therefore, weakly coupled superconductors are the best candidate to
observe quantum size effects and shape resonances in the superconducting
gaps (and in the critical temperature).

\begin{table}
\caption{Bulk values of the gap normalized to $\omega_0$ and the amplification factors $A$ at $L=0.8$ nm for the different couplings $\lambda$ here considered.}
\label{tab_1}
\begin{center}
\begin{tabular}{lll}
\hline\noalign{\smallskip}
$\lambda$ & $\frac{\Delta_{bulk}}{\omega_0}$ &A=$\frac{\Delta_{max}}{\Delta_{bulk}}$  \\
\noalign{\smallskip}\hline\noalign{\smallskip}
0.3 \,\,\,  & 0.073 \,\,\, & \,\,\, 2.98 \\
0.6 \,\,\,  & 0.394 \,\,\, & \,\,\, 1.85 \\
1.0 \,\,\,  & 0.855 \,\,\, & \,\,\, 1.61 \\
1.5 \,\,\,  & 1.399 \,\,\, & \,\,\, 1.54 \\
2.0 \,\,\,  & 1.926 \,\,\, & \,\,\, 1.50 \\
\noalign{\smallskip}\hline
\end{tabular}
\end{center}
\end{table}

Note that increasing the thickness $L$ the amplification $A$
becomes less dependent on the coupling. Interestingly for experimental
detection, even for the large thickness $L=5$ nm, the amplification of the
gap is approximately 1.25, which is a measurable effect in all 
practical cases.

In Figure 2 we show the superconducting gaps in the first subband and in the
last subband contributing to the pairing
as a function of the nanofilm thickness tuned around a shape resonance
($N_{res}=10$) for different values of the energy cutoff $\omega_0$,
at fixed coupling strength
chosen in the intermediate coupling regime ($\lambda=0.6$).
As in Figure 1, the gaps are normalized to their bulk value, obtained
in the limit of large thickness ($k_FL >>1$), see Table \ref{tab_1}.
As one can see, the multigap regime of the superconducting condensate is
present only in the shape resonant region, and for the here-considered 
cases we have all the gaps equal ($\Delta_1=\Delta_2=...=\Delta_9$), 
except the gap of the last subband ($\Delta_{10}$).
For $\omega_0=300$ K, the largest difference between these gaps is found
at the anti-resonance, with a factor 1.05 of difference for the resonance
$N_{res}=10$ at $L=4.56$ nm, and the width of the resonance having multigap
character is found over width-span $\delta L=0.07$ nm.
Increasing the energy cutoff to $\omega_0=1500$ K, the width-span of the
resonance showing multiple gaps increases to $\delta L=0.37$ nm, which is
now a range of thicknesses realizable in current nanofilm deposition
processes. It is therefore crucial to consider
systems with large energy cutoffs and in the weak coupling regime
to amplify in size and in width the multigap resonant character of
the confined superconductors, in order to be able to access experimentally
the interesting multigap regime, never observed in the single superconducting 
nanofilms.

\begin{figure}
\includegraphics[scale=0.6]{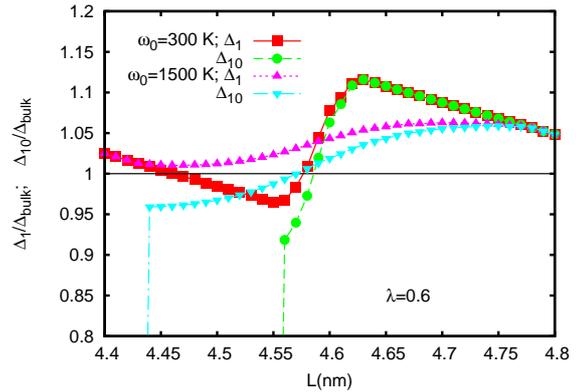}
\caption{$\Delta_1$ and $\Delta_{10}$ as a function of thickness, close to a shape resonance, and normalized to $\Delta_{bulk}$. The aim is to study the width of the shape resonance as a function of the cutoff energy and multigap structure of the condensate.}
\label{fig_2}
\end{figure}

In Figure 3 the chemical potential as a function of thickness
for different couplings and two different cutoff energies is reported.
The chemical potential $\mu$ is normalized with respect to the Fermi energy
of the 3D bulk non interacting system $E_F$, value that is approached
in the limit $(k_FL)>>1$. Since we work at fixed conduction electron
density, the chemical potential is renormalized by the discrete structure
of the electronic levels and by the superconducting gap opening.
The main effect is the discreteness of the levels, while the gap opening, both
in value and in energy extension ($2\omega_0$), determines differences
only around the shape resonant region, differences which become sizable
when $\omega_0$ and the gaps increases in the strong coupling regime.
We find that in the ultrathin limit $L<3$ nm, the solution of the coupled
gaps and density equations is important and it is not possible to work at
fixed chemical potential to get the precise locations in $L$ 
of the shape resonances.

\begin{figure}
\includegraphics[scale=0.6]{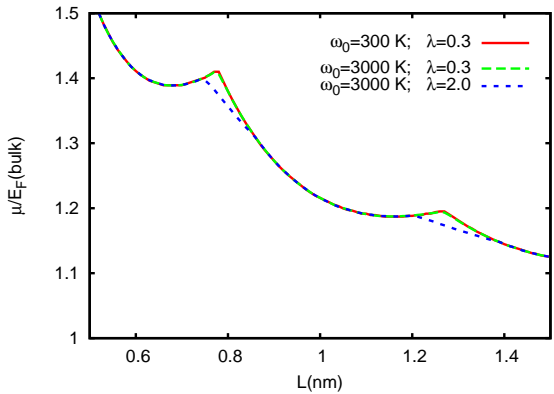}
\caption{Chemical potential as a function of thickness for different couplings and different $\omega_0$ at fixed density.}
\label{fig_3}
\end{figure}

Figure 4 shows the total gap profile $\Delta(z)$ along the 
direction transverse to the film ($z$), 
evaluated according to Eq. (\ref{gap_z}). 
For the case $\lambda=0.6$ and $\omega_0=300$ K we consider three cases: 
the thickness $L=2.210$ nm at the $N=4$ anti-resonance, $L=2.222$ nm very
close to the shape resonance, and $L=2.270$ nm outside and above the shape 
resonance.

\begin{figure}
\includegraphics[scale=0.6]{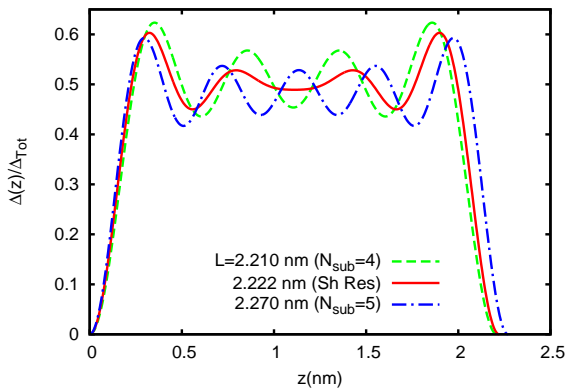}
\caption{Superconducting gap profile along the transverse direction $z$ for $\lambda=0.6$ and $\omega_0=300$ K. $\Delta(z)$ is normalized by $\Delta_{Tot}$, the integral over $z$ of the gap profile itself. Three different thicknesses are considered, before, at, and after the $N_{sub}=(4,5)$ shape resonance.}
\label{fig_4}
\end{figure}

The results shown in Figure 4 indicate an interesting behavior of the
superconducting gap profile: $\Delta(z)$ outside resonance displays
the Friedel oscillations, as already discussed in Ref. \cite{Altomarebook}, 
together with the vanishing of the gap profile at the boundaries due
to infinite-wall boundary conditions. 
The new interesting property reported here is a quite flat
behavior of $\Delta(z)$ when the nanofilm thickness $L$ is tuned very close
to the shape resonance, see the case $L=2.222$ nm in Figure 4. 
We have also analyzed other shape resonances for larger $L$, finding
an even flatter behavior at resonance, owing to the larger number
of harmonics entering in the calculation of $\Delta(z)$. 
Regarding the electron density profile $n_e(z)$ of
Eq. (\ref{density_z}), we have found a very flat dependence of $n_e(z)$
at the center of the nanofilm, with tiny oscillations as a function
of $z$ (less than $10\%$ of the maximal density).
Therefore, close to shape resonances the superconducting ground state
of the nanofilms in the quantum-size regime 
appears to be quite uniform, with the exception of the boundaries 
(for more realistic boundary conditions, see \cite{Valentinis1}),
together with sizable amplifications of the gaps, and hence it points
toward an optimized shape-resonant superconductivity.

\section{Conclusions}
In this paper we have shown that the ground state properties of
shape-resonant superconductivity in ultranarrow nanofilms strongly depend 
on the microscopic details of the pairing interaction. The amplification
of the superconducting gap is the largest in the ultrathin limit and in the
weak-coupling regime of pairing. The same amplification is progressively
reduced when the coupling is increased toward strong coupling. The width
of the shape resonance is instead governed by the energy cutoff of the
pairing interaction: the range of thicknesses of the nanofilms 
in which superconductivity is shape-resonant increases
for increasing energy cutoff, allowing the formation of a multicondensate
and multigap superconducting phase in the shape-resonant region.
Interestingly, the gap profile along the transverse direction of the nanofilm
indicates a uniform and robust superconducting state at resonances. The
multigap properties at resonance may be detected by next generation
nano-ARPES \cite{Saini2016,Vignaud2016} or nano-STM 
measurements \cite{Renner2016}, which are in construction to
investigate structural and electronic complexity in high-T$_c$ superconductors.
Therefore, we conclude that the optimal shape resonant superconductors
can be realized starting from intermediate to weak-coupling bulk
superconductors having large energy cutoffs, as in FeSe monolayers or
doped graphene systems, reducing one or more dimensions to the nano
or atomic scale.

\label{conclusions}

\begin{acknowledgements}
We acknowledge D. Valentinis, D. Van der Marel, and C. Berthod for useful discussions. A. Ricci is also acknowledged for his comments on the experimental detection of the predictions of this paper.
M. Cariglia acknowledges CNPq support from project (207007 / 2014-4) and FAPEMIG support from project APQ-02164-14. 
M.M. Doria acknowledges CNPq support from funding 
(23079.014992 / 2015-39).
M.V. Milo\v{s}evi\'c acknowledges support
from Research Foundation - Flanders (FWO).
A. Perali acknowledges financial support from
the University of Camerino under the project FAR ``Control and enhancement of superconductivity by engineering materials at the nanoscale''.
All authors acknowledge the collaboration within the MultiSuper International 
Network (http://www.multisuper.org) for exchange of ideas and suggestions.
\end{acknowledgements}

\end{document}